\documentclass[amsmath,amssymb,superscriptaddress,nobalancelastpage,prb,twocolumn]{revtex4-2} 

\AtBeginDocument{%
\newwrite\bibnotes
\def\bibnotesext{Notes.bib}
\immediate\openout\bibnotes=\jobname\bibnotesext
\immediate\write\bibnotes{@CONTROL{REVTEX41Control}}
\immediate\write\bibnotes{@CONTROL{%
apsrev41Control,author="08",editor="1",pages="1",title="0",year="1"}}
\if@filesw
\immediate\write\@auxout{\string\citation{apsrev41Control}}%
\fi
}%

\usepackage{graphicx}
\usepackage{varioref}
\usepackage{xr-hyper}
\usepackage{xcolor}
\usepackage{hyperref}
\usepackage{ulem}
\usepackage{braket}
\usepackage{varioref}
\usepackage{xr-hyper}
\usepackage{xcolor}
\usepackage{hyperref}
\hypersetup{colorlinks,linkcolor=blue,urlcolor=blue,citecolor=blue}
\newcommand{\Nd}{Nd-LSCO ($p=0.20$, $T_c=20$~K)}
\newcommand{\Eu}{Eu-LSCO ($p=0.21$, $T_c=15$~K)}

\begin{document}


\title{Pseudogap Suppression by Competition with Superconductivity in La-Based Cuprates}

 \author{J. K\"uspert}
        \affiliation{Physik-Institut, Universit\"{a}t Z\"{u}rich, Winterthurerstrasse 190, CH-8057 Z\"{u}rich, Switzerland}
        
         \author{R. Cohn Wagner}
        \affiliation{Physik-Institut, Universit\"{a}t Z\"{u}rich, Winterthurerstrasse 190, CH-8057 Z\"{u}rich, Switzerland}
        
         \author{C. Lin}
        \affiliation{Physik-Institut, Universit\"{a}t Z\"{u}rich, Winterthurerstrasse 190, CH-8057 Z\"{u}rich, Switzerland}
        
         \author{K. von Arx}
        \affiliation{Physik-Institut, Universit\"{a}t Z\"{u}rich, Winterthurerstrasse 190, CH-8057 Z\"{u}rich, Switzerland}
        \affiliation{Department of Physics, Chalmers University of Technology, SE-412 96 G\"oteborg, Sweden}
        
           \author{Q. Wang}
        \affiliation{Physik-Institut, Universit\"{a}t Z\"{u}rich, Winterthurerstrasse 190, CH-8057 Z\"{u}rich, Switzerland}
        
         \author{K. Kramer}
        \affiliation{Physik-Institut, Universit\"{a}t Z\"{u}rich, Winterthurerstrasse 190, CH-8057 Z\"{u}rich, Switzerland}

   \author{W. R. Pudelko}
        \affiliation{Physik-Institut, Universit\"{a}t Z\"{u}rich, Winterthurerstrasse 190, CH-8057 Z\"{u}rich, Switzerland}
        \affiliation{Swiss Light Source, Paul Scherrer Institut, CH-5232 Villigen PSI, Switzerland}
        
     \author{N.~C.~Plumb}
        \affiliation{Swiss Light Source, Paul Scherrer Institut, CH-5232 Villigen PSI, Switzerland}
        
    \author{C. E.~Matt}
        \affiliation{Physik-Institut, Universit\"{a}t Z\"{u}rich, Winterthurerstrasse 190, CH-8057 Z\"{u}rich, Switzerland}
        \affiliation{Swiss Light Source, Paul Scherrer Institut, CH-5232 Villigen PSI, Switzerland}
  
    \author{C. G. Fatuzzo}
        \affiliation{Institute of Physics, \'{E}cole Polytechnique Fed\'{e}rale de Lausanne (EPFL), CH-1015 Lausanne, Switzerland}
  
    \author{D.~Sutter}
        \affiliation{Physik-Institut, Universit\"{a}t Z\"{u}rich, Winterthurerstrasse 190, CH-8057 Z\"{u}rich, Switzerland}

    \author{Y.~Sassa}
        \affiliation{Department of Physics, Chalmers University of Technology, SE-412 96 G\"oteborg, Sweden}
        
        \author{J. -Q. Yan}
\affiliation{Materials Science and Technology Division, Oak Ridge National Laboratory, Oak Ridge, Tennessee 37831, United States}
\author{J. -S. Zhou} 
\affiliation{Texas Materials Institute, University of Texas at Austin, Austin, Texas 78712, USA} 
\author{J. B. Goodenough}
\affiliation{Texas Materials Institute, University of Texas at Austin, Austin, Texas 78712, USA} 

        \author{S.~Pyon}
\affiliation{Department of Advanced Materials, University of Tokyo, Kashiwa 277-8561, Japan}
\author{T.~Takayama}
\affiliation{Department of Advanced Materials, University of Tokyo, Kashiwa 277-8561, Japan}
\author{H.~Takagi}
\affiliation{Department of Advanced Materials, University of Tokyo, Kashiwa 277-8561, Japan}
       
  \author{T.~Kurosawa}
\affiliation{Department of Physics, Hokkaido University - Sapporo 060-0810, 
Japan}
 
\author{N.~Momono}
\affiliation{Department of Physics, Hokkaido University - Sapporo 060-0810, 
Japan}
\affiliation{Department of Applied Sciences, Muroran Institute of Technology, 
Muroran 050-8585, Japan}

\author{M.~Oda}
\affiliation{Department of Physics, Hokkaido University - Sapporo 060-0810, 
Japan}

    \author{M.~Hoesch}
    \affiliation{Diamond Light Source, Harwell Campus, Didcot, OX11 0DE, United Kingdom}
    \affiliation{DESY Photon Science, Notkestraße 85, 22607 Hamburg, Germany}
        
     \author{C. Cacho}
    \affiliation{Diamond Light Source, Harwell Campus, Didcot, OX11 0DE, United Kingdom}

    \author{T.~K.~Kim}
    \affiliation{Diamond Light Source, Harwell Campus, Didcot, OX11 0DE, United Kingdom}
          
\author{M.~Horio}
        \affiliation{Physik-Institut, Universit\"{a}t Z\"{u}rich, Winterthurerstrasse 190, CH-8057 Z\"{u}rich, Switzerland}
 
    \author{J.~Chang}
    \affiliation{Physik-Institut, Universit\"{a}t Z\"{u}rich, Winterthurerstrasse 190, CH-8057 Z\"{u}rich, Switzerland}

\begin{abstract}
We have carried out a comprehensive high-resolution angle-resolved photoemission spectroscopy (ARPES) study of the pseudogap interplay with superconductivity in La-based cuprates. The three systems La$_{2-x}$Sr$_x$CuO$_4$, La$_{1.6-x}$Nd$_{0.4}$Sr$_x$CuO$_4$, and La$_{1.8-x}$Eu$_{0.2}$Sr$_x$CuO$_4$ display slightly different pseudogap critical points in the temperature versus doping phase diagram.
We have studied the pseudogap evolution into the superconducting state for doping concentrations just below the critical point. In this setting, near optimal doping for superconductivity and in the presence of the weakest possible pseudogap, we uncover how the pseudogap is partially suppressed inside the superconducting state. This conclusion is based on  the direct observation of a reduced pseudogap energy scale and re-emergence of spectral weight suppressed by the pseudogap. Altogether these observations suggest that the pseudogap phenomenon in La-based cuprates is in competition with superconductivity for anti-nodal spectral weight.  
\end{abstract}

\maketitle
  
\section{Introduction}

Strange metal behavior~\cite{LegrosNatPhys2018} and 
pseudogap physics~\cite{normanAP} remain the most challenging problems of the cuprate superconductors. One 
key characteristic of strange metals is that resistivity scales uninterrupted with thermal excitation energy down to the lowest measurable temperature. In the cuprates, this is observed at a critical doping $p^*$~\cite{daou09,CooperScience2009}.
Above $p^*$, standard Fermi liquid properties are restored~\cite{NakamaePRB2003}, whereas below it, a mysterious pseudogap phenomenon emerges. The pseudogap manifests in multiple experiments. In spectroscopic measurements, the pseudogap is associated with a partial gaping of spectral weight near the Fermi level~\cite{HashimotoNatPhys14}. Numerous studies have attempted to address the nature of the pseudogap~\cite{normanAP}. These experiments are typically carried out in the normal state, aiming to connect the pseudogap to either superconducting fluctuations~\cite{WangPRB2006}, a symmetry breaking order parameter~\cite{daou10,FauquePRL2006,HashimotoNatPhys2010}, or a cross-over phenomenon~\cite{TallonPRB2020}. 
There is much less experimental work addressing the pseudogap inside the superconducting state~\cite{ShekhterNature2013}. In the very underdoped regime, photoemission studies point to a competition between the pseudogap phenomenon and superconductivity -- with the latter being partially suppressed by the former~\cite{kondo}. It has, however, been difficult to tune or influence the pseudogap, which appears rather insensitive to disorder or magnetic field \cite{RullierAlbenque2006, RullierAlbenque2008}. Theoretical and experimental work suggests that the pseudogap critical point $p^*$ is confined by the van Hove singularity crossing of the Fermi level~\cite{BenhabibPRL2015,WuPRX2018}. 

Tuning the van Hove singularity by hydrostatic pressure is one way to manipulate the pseudogap phenomenon~\cite{DoironLeyraudNatComm2017}. Another route is to identify interactions with other phases. The pseudogap has been shown  to suppress the superconducting order parameter~\cite{kondo}. Much less is known about the reciprocal relation, namely how superconductivity influences the pseudogap phenomenon. An unsolved problem relates to the interplay between the pseudogap and superconductivity in the regime close to $p^*$. This issue has been difficult to address since it challenges both temperature and energy resolution limitations of most synchrotron angle-resolved photoemission spectroscopy (ARPES)~\cite{sobota_angle-resolved_2021} beamlines.



\begin{figure*}
	\begin{center}
		\includegraphics[width=0.995\textwidth]{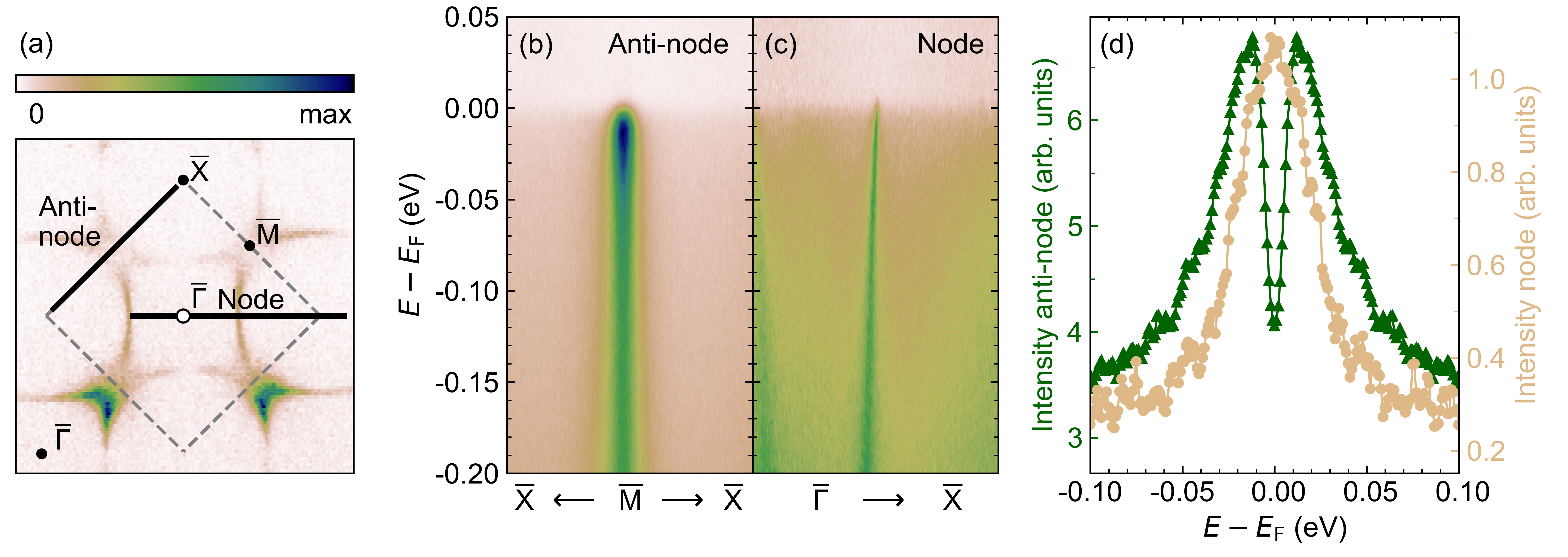}
	\end{center}
 	\caption{Photoemission intensities recorded on LSCO $p=0.145$ in the superconducting state. (a) Fermi surface map, recorded with $h\nu =$ 160\,eV photons at $T =$ 7\,K, and  integrated $\pm$16\,meV around the Fermi level. The Brillouin zone boundary is indicated by the dashed grey lines and high symmetry points are labeled $\overline{\Gamma}$, $\overline{\text{X}}$, and $\overline{\text{M}}$. Solid black lines indicate nodal and anti-nodal directions.
 	(b) and (c) ARPES spectra recorded at $T =$ 6\,K along the anti-nodal ($\overline{\text{M}} \rightarrow \overline{X}$) and nodal ($\overline{\Gamma} \rightarrow \overline{\text{X}}$) directions using $h\nu =$ 55\,eV photons.
 	(d) Nodal (beige circles) and anti-nodal (green triangles) symmetrized energy distribution curves (EDCs) at the Fermi momentum, after background subtraction as described in Ref.\,\cite{Mattnatcom2018}.}
	\label{fig1}
\end{figure*}  

Here, we study the pseudogap in the limit $p\rightarrow p^*$, where $T^*$ approaches $T_\text{c}$. 
We  chose to examine La-based cuprates, in which the pseudogap energy scale is much larger than the superconducting gap amplitude~\cite{MattPRB2015, sobota_angle-resolved_2021}. Investigating these compounds, where $\Delta^*\gg \Delta_{sc}$, enabled us to track both, the 'pure' pseudogap as well as its interplay with superconductivity. Using the low-temperature and high energy resolution capabilities of the I05 beamline at 
Diamond Light Source, 
we have explored the evolution of the pseudogap inside the superconducting state of La$_{2-x}$Sr$_x$CuO$_4$ (LSCO) $x=p=0.145$,  La$_{1.8-x}$Eu$_{0.2}$Sr$_x$CuO$_4$ (Eu-LSCO) $p=0.21$ and La$_{1.6-x}$Nd$_{0.4}$Sr$_x$CuO$_4$ (Nd-LSCO) $p=0.20$. We find that the pseudogap amplitude is partially suppressed inside the superconducting state,
suggesting a competing interaction.
As a defining property of the pseudogap phase, we observe an anti-nodal spectral weight suppression for $T<T^*$. Below $T^*$, we identify a third temperature scale $T^\dagger>T_\text{c}$, below which anti-nodal weight partially recovers. Eventually, complete recovery is found for $T\rightarrow 0$. This spectral weight recovery is discussed in the context of a tri-phase competition between superconductivity, charge order, and pseudogap physics.

\section{Methods} 
Single crystals of LSCO ($p=0.145$, $T_\text{c}=37$\,K ~\cite{ming,ChangPRL09} and $p=0.12$, $T_\text{c}=27$\,K~\cite{ChangPRB08}), \Nd~\cite{MattPRB2015} and \Eu\ were synthesized by the traveling floating zone method. 
The critical pseudogap dopings are $p^* \approx 0.19$~\cite{CooperScience2009} for LSCO and $p^* \approx 0.23$~\cite{CollignonPRB2017, Michon2019} for Nd-LSCO and Eu-LSCO. ARPES experiments were carried out at  beamline I05~\cite{HoeschRevSciInst2017_full} of Diamond Light Source and the Surface and Interface Spectroscopy (SIS) beamline of Swiss Light Source. Single crystals were mechanically cleaved {\it in-situ} in ultra-high vacuum using top posts. Measurements were performed using 55\,eV or 160\,eV linear-horizontally polarized light at I05 and circularly polarized light at SIS. Dependent on photon energy and instrument, the energy resolution (Gaussian standard deviation) spans in the range of 5\,-\,15\,meV.

    \begin{figure*}
	\begin{center}
		\includegraphics[width=\textwidth]{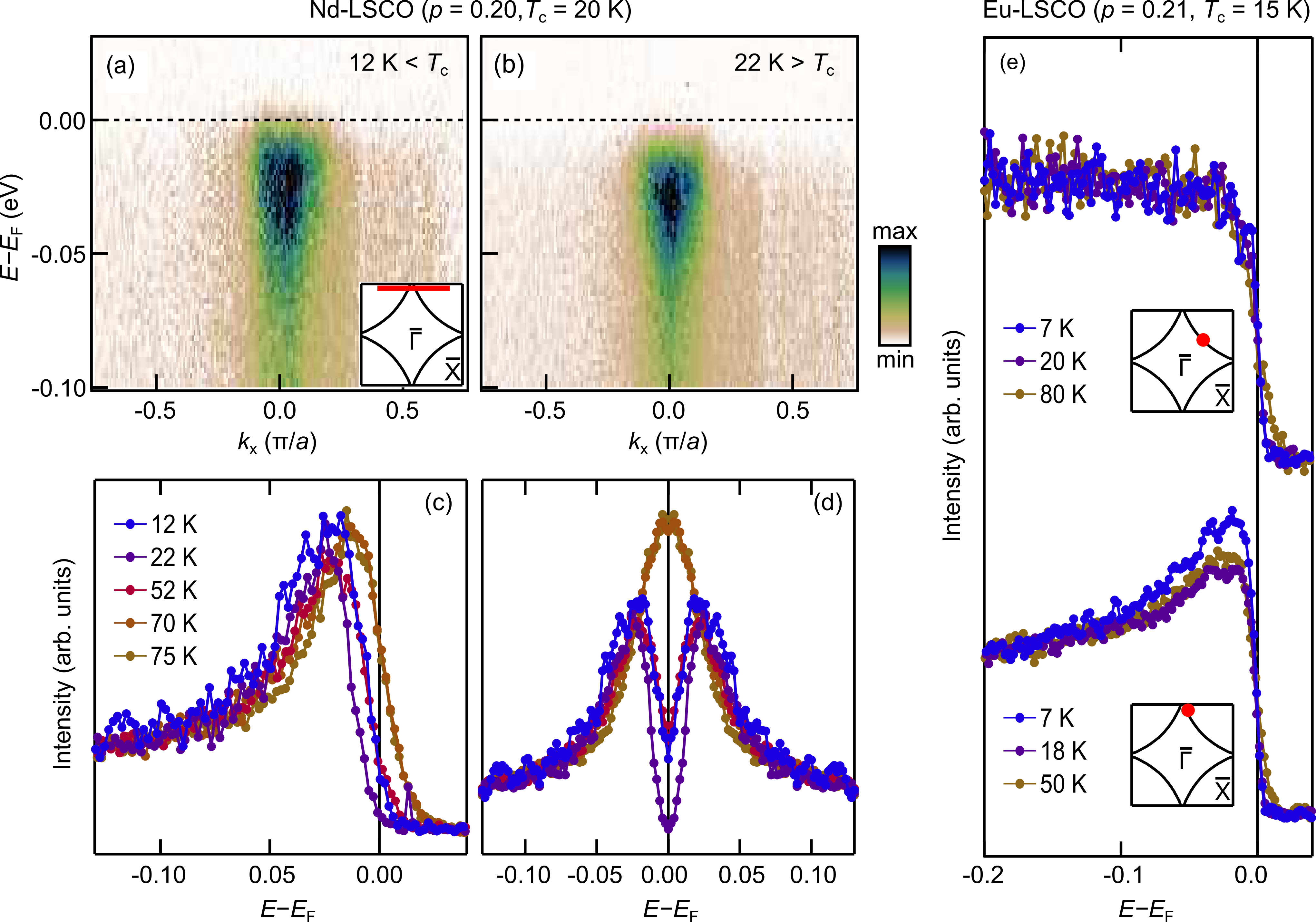}
	\end{center}
 	\caption{Temperature dependence of nodal and anti-nodal spectra in Nd-LSCO ($p=0.20$) and Eu-LSCO ($p=0.21$) at $h\nu = 55$\,eV. (a,b) Energy distribution maps along the anti-nodal direction (see inset), recorded on Nd-LSCO \cite{MattPRB2015} for temperatures as indicated. (c,d) Corresponding energy distribution curves (EDCs) and symmetrized EDCs at the underlying Fermi momentum. (e) Temperature-dependent nodal and anti-nodal spectra recorded on Eu-LSCO. Pseudogap temperatures for Nd-LSCO $p=0.20$ and Eu-LSCO $p=0.21$ are $T^* \approx 80$\,K and  $T^* \approx 75$\,K~\cite{MattPRB2015,OlivierPRB2018}.}
	\label{fig2}
\end{figure*}
    
\section{Results}  
We studied three different La-based compounds  (LSCO, Nd-LSCO, and Eu-LSCO). 
Consistent with existing ARPES literature, the data quality obtained from LSCO and Nd-LSCO crystals~\cite{MattPRB2015, ChangNJP2008} is comparatively better than that recorded on Eu-LSCO~\cite{ZabolotnyyEPL2009}. In Fig.\,\ref{fig1}(a)-(c), we display a Fermi surface map, nodal and anti-nodal spectra recorded on LSCO $p=0.145$.
 Symmetrized~\cite{normanNAT98} nodal and anti-nodal energy distribution curves (EDCs) at $k_\text{F}$ are shown in Fig.\,\ref{fig1}(d).
 These results are directly comparable 
 to a previous study of this compound~\cite{ming}. The improved data quality stems from higher energy resolution and smaller beam spot. These advances result in a higher signal-to-background ratio that 
 we exploit to  study the pseudogap phenomenon. Anti-nodal spectra recorded on Nd-LSCO $p=0.20$ are depicted in Fig.~2 for temperatures as indicated. The pseudogap spectra $T_\text{c}<T<T^*$ have been discussed in a previous publication~\cite{MattPRB2015}.  Here, we also enter the superconducting state. From the raw energy distribution maps of Nd-LSCO shown in Fig.\,\ref{fig2}(a, b), it is directly visible that the spectral gap at 22\,K, just above $T_\text{c}$, is larger than that inside the superconducting state. This is further confirmed by analyzing the EDCs at the underlying Fermi momentum $k_\text{F}$ -- see Fig.\,\ref{fig2}(c, d). The symmetrized anti-nodal EDCs display an effectively smaller gap inside the superconducting state than what is observed for $T\approx T_\text{c}$.  
The results on Eu-LSCO $p=0.21$ 
reveal the increase of anti-nodal spectral weight inside the superconducting state while no such gain is detectable at the nodal point -- see Fig.\,\ref{fig2}(e). Both observations suggest
a weakening of the pseudogap that is in general characterized by increasing suppression of spectral weight as the temperature is lowered.
 
   \begin{figure*}
	\begin{center}
		\includegraphics[width=0.995\textwidth]{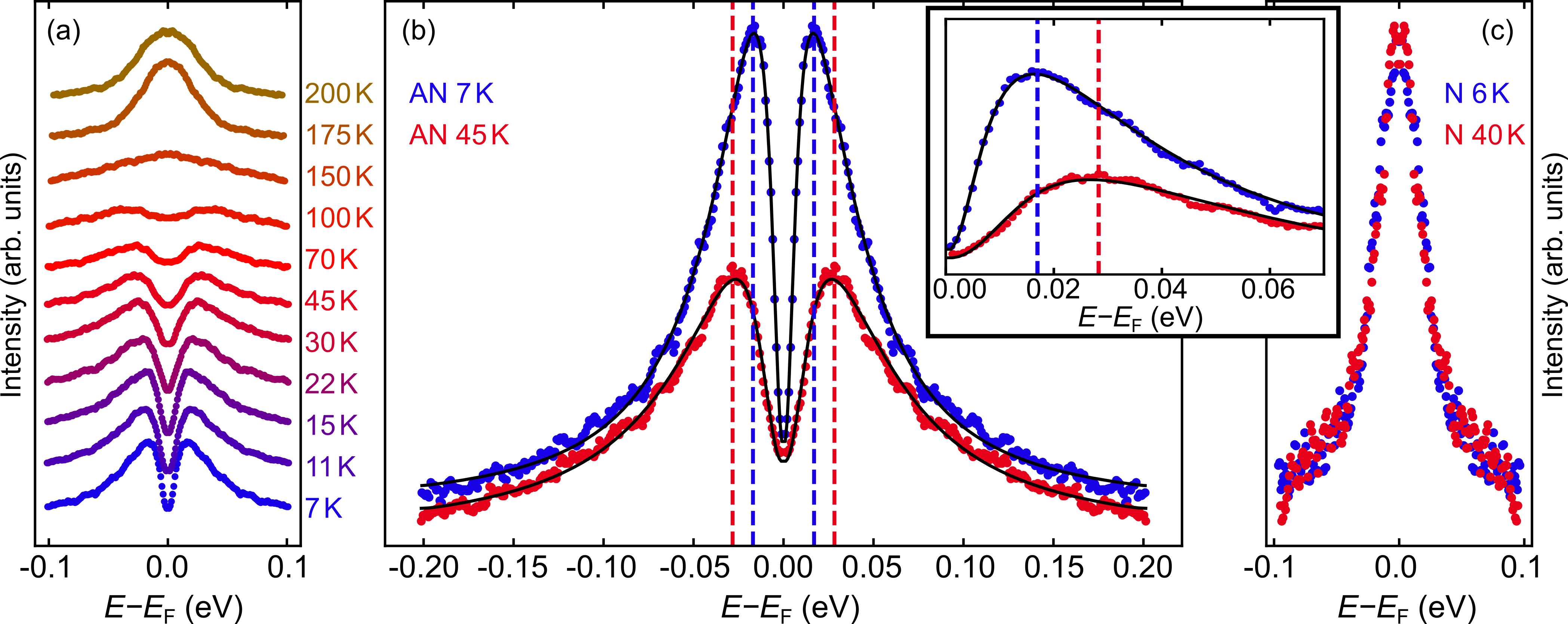}
	\end{center}
 	\caption{Nodal and anti-nodal spectra versus temperature in LSCO $p=0.145$ ($T_\text{c}=37$\,K) at $h\nu = 55$\,eV. (a) Symmetrized anti-nodal spectra for temperatures as indicated. The pseudogap onset temperature of $T^*\approx 162.5$~K, deduced from spectral weight analysis of the EDCs, is consistent with that extracted from transport measurements~\cite{OlivierPRB2018}. (b, c) Comparison of symmetrized anti-nodal and nodal spectra for $T\ll T_\text{c}$ (blue) and $T \gtrsim T_\text{c}$ (red). Solid black lines are fits using a  phenomenological self-energy function~\cite{NormanPRB1998} -- see text.
 	Vertical dashed lines indicate the peak position. The inset displays a zoom of the low-energy part of the symmetrized EDCs in (b). A background defined using the methodology given in Ref.\,\cite{Mattnatcom2018} has been subtracted from all spectra.}
	\label{fig3}
\end{figure*}
 
 The Nd-LSCO and Eu-LSCO compounds (space group 138~\cite{HuckerPRB2011}) are special because they have additional chemical disorder due to substitution of neodymium and europium. This substitution stabilizes the so-called low-temperature tetragonal phase. 
 We therefore 
 additionally investigated the LSCO $p=0.145$ compound that has less chemical disorder and a different crystal structure (space group 64~\cite{RadaelliPRB94, BRADEN1992, Frison2022}). 
 The larger $T_\text{c}$ of this compound allowed to probe deep into the superconducting state.
Background subtracted~\cite{Mattnatcom2018} anti-nodal EDCs are shown in Fig.\,\ref{fig3}(a) as a function of temperature. The pseudogap opens at $T^*\approx 162.5$\,K. As in all other hole-doped cuprates, the pseudogap manifests itself by a loss of spectral weight near the Fermi level. Upon cooling, the weight loss gradually increases. In Fig.\,\ref{fig3}(b), the anti-nodal spectrum at $T=45$~K, just above the superconducting transition, is shown.  We find a pronounced re-emergence of anti-nodal spectral weight inside the superconducting state, as exemplarily shown by the spectrum taken at $T=7$\,K (see Fig.\,\ref{fig3}(b)). This is in strong contrast to the nodal spectra, which are essentially temperature independent (Fig.\,\ref{fig3}(c)). 

Complementary to our observation of spectral weight loss, we find a peculiar temperature dependence of the pseudogap. The vertical dashed lines in Fig.\,\ref{fig3}(b) indicate the peak positions in the symmetrized anti-nodal EDCs. Defining the gap amplitude by half the distance between the peaks yields a reduction of the gap amplitude for $T\ll T_\text{c}$.

 \section{Analysis}
\begin{figure*}{}
	\begin{center}
		\includegraphics[width=1\textwidth]{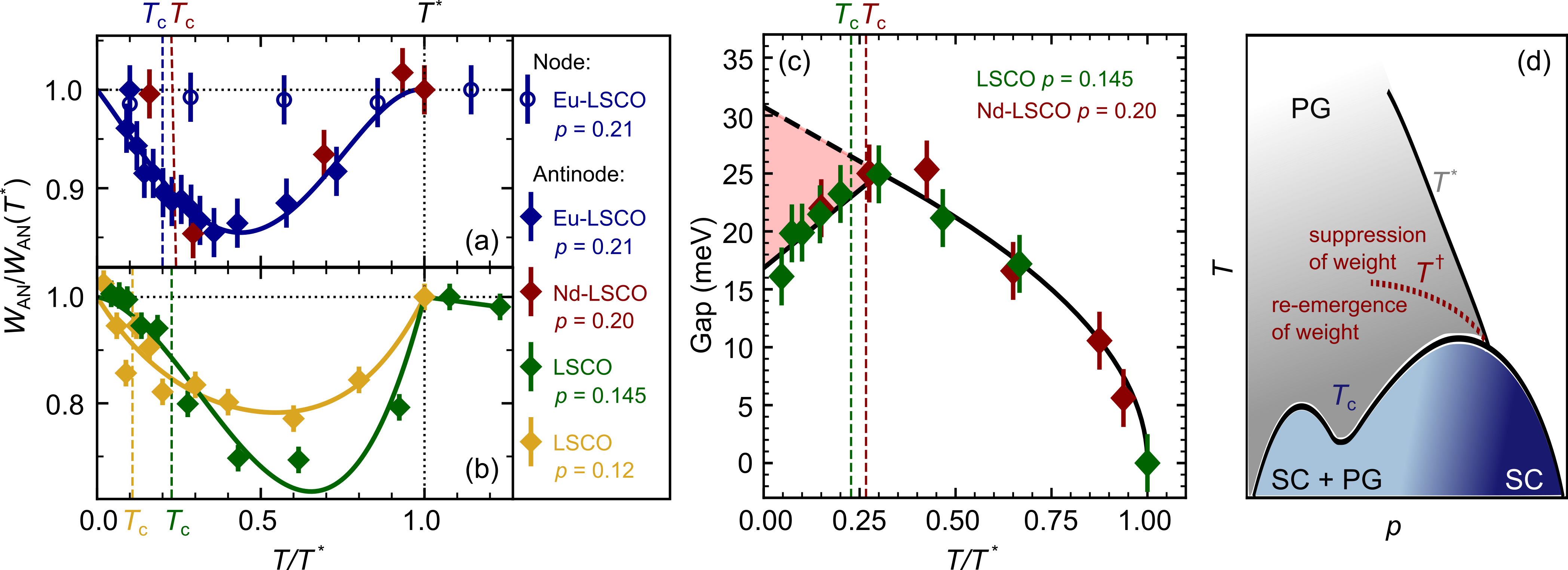}
	\end{center}
 	\caption{Anti-nodal competition between the pseudogap and superconductivity. (a), (b) Integrated spectral weight as a function of temperature. 
 	Anti-nodal weights of LSCO, Eu-LSCO, and Nd-LSCO are plotted with triangles whereas open circles denote nodal weight. Solid lines are guides to the eye. Vertical dashed lines indicate $T=T^*$ and $T_\text{c}/T^*$, respectively. Error bars provide an estimate of the systematic uncertainty. 
 	(c) Anti-nodal spectral gap as a function of temperature 
 	for LSCO (this work) and Nd-LSCO~\cite{MattPRB2015}. The gap amplitude follows a linear dependency for $T \lesssim T_\text{c}$ and can be fitted by an order parameter like $(1-T/T^*)^{0.5}$ behavior for $T \gtrsim T_\text{c}$. The energy resolution defined by the standard deviation of the
Fermi step sets the error bars.
 	(d) Phase diagram (temperature versus doping) indicating phase space of different anti-nodal spectral weight behavior. Outside the pseudogap phase, spectral weight is conserved. We show that there exists a temperature scale $T^\dagger<T^*$ below which spectral weight is partially recovered.}
	\label{fig4}
\end{figure*}

  \textit{Spectral weight:} 
  A defining property of the pseudogap is the partial suppression of anti-nodal spectral weight 
  $I(k_{AN}, \omega$). We define the integrated spectral weight as $W_{i}=\int d\omega [I(k_i,\omega)+I(k_i,-\omega)]$ 
  with $i={AN,N}$ being anti-nodal or nodal. Our integration window of the symmetrized EDCs is 
  $-0.2<\omega<0.2$~eV.
  The nodal spectral weight $W_N$ is essentially temperature independent (see circles in Fig.\,\ref{fig4}(a) and Fig.\ref{fig2}(e), \ref{fig3}(c)). In contrast, anti-nodal spectral weights display a significant temperature dependence, as shown by the triangles in Fig.\,\ref{fig4}(a, b).
  When entering the pseudogap state, $W_{AN}$ is suppressed and gradually diminishes upon cooling. However, for Eu-LSCO, Nd-LSCO, and LSCO, a gradual recovery is observed below a temperature scale $T^{\dagger}$. In LSCO $p=0.12$ and $0.145$, $T^{\dagger}\sim 2-3\, T_\text{c}$ while in Eu-LSCO $p=0.21$, $T^{\dagger}$ is closer to $T_\text{c}$. The spectral weight recovery continues inside the superconducting state.  Eventually, as $T\rightarrow 0$, $W_{AN}$
  fully recovers to the level of weight observed above $T\leq T^*$.
  
  \textit{Gap analysis:} To extract the amplitude of the anti-nodal spectral gap, we employ the spectral function 
  $A(k_F,\omega)= -\pi^{-1}\Sigma''/[(\omega-\Sigma')^2+\Sigma''^2]$
  convoluted with a Gaussian distribution to mimic experimental resolution~\cite{NormanPRB1998}. $\Sigma'$ and $\Sigma''$ are the real and imaginary parts of the self-energy, respectively. Our problem involves at least a superconducting gap and a pseudogap. Since the nature of the pseudogap is not established, a microscopic understanding of its self-energy is missing. Eliashberg theory, in contrast,  describes the self-energy effect associated with superconductivity~\cite{ChubukovPRB2007}. If the pseudogap was a precursor to superconductivity, the Eliashberg framework would also apply to the pseudogap state. 
  
  There are several experimental indications that the pseudogap is not associated with superconducting fluctuations~\cite{HashimotoNatPhys14,ChangNatPhys2012}. 
  In Nd-LSCO, for example, there is strong evidence of vanishing of the pseudogap at a quantum critical point inside the superconducting dome~\cite{daou09}. In Nd-LSCO $p=0.20$,
 we observe a spectral gap at the superconducting onset $T_\text{c}=20$~K.  
 Mean-field theory yields $2\Delta=\alpha k_\text{B}T_\text{c}$ where $\alpha=4.3$~\cite{HeScience2018}
 for weakly coupled $d$-wave superconductors.
The gap amplitude of 20-25\,meV (Fig.\,\ref{fig4}(c)) implies $\alpha \sim 20$. Although the coupling coefficient $\alpha$ can be larger in the strong coupling limit, this appears unreasonably large. In contrast, assigning the gap to the pseudogap onset temperature ($2\Delta=\alpha k_\text{B}T^*$) yields a more reasonable $\alpha\approx5$. 
 We thus associate the observed spectral gap with the pseudogap phenomenon and assume that the superconducting gap is not directly detectable due to the finite energy resolution. In LSCO, the differentiation of the pseudogap and superconducting gap is less obvious. Around optimal doping, the pseudogap energy scale is smaller than that in Nd-LSCO. At the same time, the superconducting gap is expected to be larger due to its 
 larger $T_\text{c}$.Within the experimental resolution, it was not possible to differentiate these two gaps. The anti-nodal spectra of LSCO are fitted using a single gap model.
 Using the phenomenological ansatz~\cite{NormanPRB1998} $\Sigma''= -\Gamma = \textrm{constant}$ and $\Sigma'=\Delta ^ 2  / \omega$, the spectral weights of LSCO and Nd-LSCO can be parametrized. With this function, a single gap energy scale is extracted as a function of temperature (see Fig.\,\ref{fig4}(c)). 
 In the pseudogap, the gap follows roughly an order parameter like $(1-T/T^*)^{0.5}$ dependence. 
 The temperature dependence is interrupted for $T<T_\text{c}$, where the amplitudes of the gaps decrease with decreasing temperature. Thus, when the pseudogap is analyzed, either by spectral weight or gap amplitude, suppression is observed below two different temperature scales. The gap amplitude is partially suppressed upon entering the superconducting state, whereas spectral weight decreases below $T^*$. Recovery is found below a temperature scale $T^\dagger$ much larger than $T_\text{c}$.
 

\section{Discussion}
It is interesting to compare our results on La-based cuprates with previous ARPES studies in (Bi,Pb)$_2$(Sr,La)$_2$CuO$_{6+\delta}$ (Bi2201)~\cite{kondo} and Bi$_2$Sr$_2$CaCu$_2$O$_{8+\delta}$ (Bi2212)~\cite{VishikPNAS2012}. 
In all systems, the onset of the pseudogap is heralded by the suppression of $W_{AN}$ and the opening of an anti-nodal spectral gap. This conclusion holds even if slightly different definitions of integrated spectral weight are employed. Upon further cooling, $W_{AN}$ diminishes, and the pseudogap energy scale increases. However, the three systems react differently upon approaching the superconducting state. In Bi2212, the recovery of $W_{AN}$ has a sharp onset at the superconducting transition~\cite{Hashimoto_NatMater141_2015, VishikPNAS2012}. For LSCO and Eu-LSCO, the recovery of $W_{AN}$ starts already below a temperature scale $T^\dagger>T_\text{c}$  -- as schematically illustrated in Fig.\,\ref{fig4}(d). 
If the recovery is interpreted in terms of phase competition with superconductivity, it must involve superconducting fluctuations in the normal state. Although such superconducting fluctuations are known to exist~\cite{ChangNatPhys2012,WangPRB2006,LiPRB2010,OlivierPRB2018,HePRX2021}, it is not obvious that they would impact the pseudogap stronger in La-based cuprates. Another possibility is that the charge ordering~\cite{tranquada,hucker,Labiberte,FinkPRB2009} competes with the pseudogap.
Charge order is expected to generate an additional temperature and energy scale. The latter has been reported in the normal state of YBa$_2$Cu$_3$O$_{7-x}$ (YBCO) and Bi2212~\cite{Vishik_2018}.
A triphase competition~\cite{BlancoPRL2013,HuckerPRB2014} between the pseudogap, charge (stripe) order, and superconductivity is likely expressed differently in La- and Bi-based cuprates, explaining the different phenomenology in different cuprate systems.  

Inside the superconducting state, the systems also behave differently. Both Bi2201 and Bi2212 display so-called coherence peaks~\cite{He_Science3316024_2011,Fedorov_Phys.Rev.Lett.8210_1999,Wei_Phys.Rev.Lett.1019_2008}. In Bi2212, the coherence peak associated with superconductivity appears at an energy scale smaller than the pseudogap~\cite{He_Science3316024_2011}. For Bi2201, on the contrary, the two energy scales are comparable~\cite{kondo}. Certainly, for Nd-LSCO $p=0.20$, as discussed above, we expect the superconducting energy scale to be much smaller than the pseudogap. This aligns with the fact that no superconducting coherence peak is observed. It is difficult to distinguish the changes in spectral weight associated with superconductivity and the pseudogap state. Generally, superconductivity is expected to redistribute spectral weight from below to above the cooper-pairing energy scale~\cite{kondo}.
No net gain or loss of spectral weight is expected from the emergence of superconductivity. The recovery of $W_{AN}$ suggests a competing interaction between superconductivity and the pseudogap state (see Fig.\,\ref{fig4}(a, b)). This interpretation is further reinforced by the observation of a diminishing pseudogap energy scale inside the superconducting state.

\section{Conclusions}
In summary, we have carried out a high-resolution ARPES study of the pseudogap inside the superconducting state. Three La-based systems (Eu-LSCO, Nd-LSCO, and LSCO) were investigated for doping concentrations just below the critical doping $p^*$,  above which the pseudogap phenomenon vanishes. 
We observe that the total anti-nodal spectral weight is suppressed below the pseudogap temperature and begins to recover already above the superconducting onset $T_\text{c}$. The pseudogap energy scale grows with decreasing temperature until $T_\text{c}$ is reached and is partially suppressed inside the superconducting state. 
Our results are different from what has been reported in single-\cite{kondo}  and bi-layer~\cite{Hashimoto_NatMater141_2015, VishikPNAS2012}  Bi-based cuprates. We interpret this as a tri-interaction between superconductivity, charge order, and pseudogap physics that is expressed differently across material systems. \\

\textit{Acknowledgments:} J.K, C.L., K.v.A., Q.W., K.K., M.H., W.R.P, N.P, and J.C. acknowledge support from the Swiss National Science Foundation (project no. 200021\_188564 and 200021\_185037). J.K. is supported by a Ph.D. fellowship from the German Academic Scholarship Foundation. 
Y.S. was supported by the Wenner-Gren foundation. This work was performed at the SIS at the Swiss Light Source and I05 at the Diamond Light Source. We acknowledge Diamond Light Source for the time on beamline I05 under proposals SI 27768-1 and SI 10550. 

%

%

%
\end{document}